\documentclass[
 reprint,
 aps,
 pra,
 eqsecnum,
 nofootinbib,
 groupedaddress,
]{revtex4-1}

\RequirePackage{custom}

\usepackage{graphicx}
\usepackage{dcolumn}
\usepackage{bm}

\begin{document}


\title{On the Perturbation of Synchrotron Motion in the Micro-Bunching Instability}

\author{Tobias Boltz}
 \email{tboltz@slac.stanford.edu}
 \altaffiliation[now at ]{SLAC, Menlo Park, USA}
 \affiliation{Karlsruhe Institute of Technology, Kaiserstraße 12, 76131 Karlsruhe, Germany}
\author{Miriam Brosi}
 \altaffiliation[now at ]{MAX IV Laboratory, Lund, Sweden}
 \affiliation{Karlsruhe Institute of Technology, Kaiserstraße 12, 76131 Karlsruhe, Germany}
\author{Bastian Haerer}
 \affiliation{Karlsruhe Institute of Technology, Kaiserstraße 12, 76131 Karlsruhe, Germany}
\author{Patrik Schönfeldt}
 \altaffiliation[now at ]{DLR-VE, Oldenburg, Germany}
 \affiliation{Karlsruhe Institute of Technology, Kaiserstraße 12, 76131 Karlsruhe, Germany}
\author{Patrick Schreiber}
 \affiliation{Karlsruhe Institute of Technology, Kaiserstraße 12, 76131 Karlsruhe, Germany}
\author{\\Minjie Yan}
 \affiliation{Karlsruhe Institute of Technology, Kaiserstraße 12, 76131 Karlsruhe, Germany}
\author{Anke-Susanne Müller}
 \affiliation{Karlsruhe Institute of Technology, Kaiserstraße 12, 76131 Karlsruhe, Germany}

\date{\today}

\begin{abstract}
The self-interaction of short electron bunches with their own radiation field can have a significant impact on the longitudinal beam dynamics in a storage ring. While higher bunch currents increase the power of the emitted coherent synchrotron radiation (CSR) which can be provided to dedicated experiments, it simultaneously amplifies the strength of the self-interaction. Eventually, this leads to the formation of dynamically changing micro-structures within the bunch and thus fluctuating CSR emission, a phenomenon that is generally known as micro-bunching or micro-wave instability. The underlying longitudinal dynamics can be simulated by solving the Vlasov-Fokker-Planck (VFP) equation, where the CSR self-interaction can be added as a perturbation to the Hamiltonian. In this contribution, we focus on the perturbation of the synchrotron motion that is caused by introducing this additional wake field. Therefore, we adopt the perspective of a single particle and eventually comment on its implications for collective motion. We explicitly show how the shape of the parallel plates CSR wake potential breaks homogeneity in the longitudinal phase space and propose a quadrupole-like mode as potential seeding mechanism of the micro-bunching instability. Moreover, we consider synchrotron motion above the instability threshold and thereby motivate an approach to control of the occurring micro-bunching dynamics. Using dynamically adjusted RF amplitude modulations we can directly address the continuous CSR-induced perturbation at the timescale of its occurrence, which allows for substantial control over the longitudinal charge distribution. While the approach is not limited to this particular application, we demonstrate how this can significantly mitigate the micro-bunching dynamics directly above the instability threshold. The gained insights are supported and verified using the VFP solver Inovesa and put into context with measurements at the KIT storage ring KARA (Karlsruhe Research Accelerator).
\end{abstract}

\maketitle

\section{\label{sec:Introduction}Introduction}
In order to increase the emission of coherent radiation and to shorten the generated light pulses, modern synchrotron light sources are deliberately operating with short electron bunches. The Karlsruhe Research Accelerator (KARA) thus has a dedicated short-bunch mode with picosecond-long bunches, which provides emission of coherent synchrotron radiation (CSR) up to the $\si{\tera\hertz}$ frequency range. Yet, due to self-interaction with its own radiation field, the increased CSR also leads to complex longitudinal dynamics within the electron bunch. At low bunch currents, the resulting perturbation mainly causes a slight deformation of the still fairly stationary electron distribution. However, above a particular threshold current $I_{\mathrm{th}}$ which depends on the specific machine settings of the accelerator, it leads to the formation of dynamically changing micro-structures within the bunch. As the longitudinal charge distribution varies over time, this in turn results in major fluctuations of the emitted CSR power. Such fluctuations have been measured at a wide range of facilities, e.g. \cite{byr2002,mue2011,abo2002,bil2016,shi2012,kar2010,and2000,wan2006,fei2011,car2002,
eva2012,wal1997,moc2006}. The phenomenon is generally referred to as micro-bunching or micro-wave instability. The underlying longitudinal dynamics can be simulated to high qualitative agreement by numerically solving the Vlasov-Fokker-Planck equation (VFP)~\cite{war2000,ban2010,bro2016,sch2017,ste2018}.

In this contribution we focus on the perturbation of the synchrotron motion that is caused by the CSR wake potential. Therefore, we adopt the perspective of a single particle and consider its motion in the absence of collective effects where the dynamics can be modeled by a simple one-dimensional harmonic oscillator. By introducing CSR self-interaction as a perturbation (considering the wake potential of a stationary charge distribution), the dynamics below the threshold current can easily be illustrated in the single particle picture. We will show how this perturbation breaks homogeneity in the longitudinal phase space and leads to the formation of a quadrupole-like modulation of the charge density, potentially acting as the seeding mechanism for the micro-bunching instability. Furthermore, we consider single particle motion above the instability threshold where the charge density is continuously varying in time. Here, the single particle dynamics are largely driven by a modulation of the effective potential's slope caused by the dynamic variation of the CSR wake potential. Eventually, we put these findings into context with measurements taken at the KIT storage ring KARA and, based on the gained insights, motivate an approach to control of the micro-bunching dynamics using a dedicated RF modulation scheme.

\section{\label{sec:LongitudinalBeamDynamics}Longitudinal Beam Dynamics}
As the CSR self-interaction predominantly affects the longitudinal motion of particles, the micro-bunching instability can be described in good approximation by modeling only the underlying longitudinal beam dynamics~\cite{war2000,ban2010,bro2016,sch2017,ste2018}. Due to the large number of particles within a bunch (order of $\num{e9}$ and higher), collective effects such as CSR self-interaction are conveniently described by modeling the charge distribution as a line charge and using a charge distribution function $\psi (z,E)$ over the longitudinal position $z$ and the particle energy $E$ instead of considering individual particles. The temporal evolution of this charge distribution $\psi (z,E,t)$ is governed by the VFP equation, which is introduced in the following subsection. In order to gain a better understanding of the synchrotron motion of individual particles within this distribution and the effects of the perturbation by the CSR wake potential, single particle motion is considered directly afterwards.

\subsection{\label{subsec:VFPEquation}Vlasov-Fokker-Planck Equation}
Following the treatment and notation in~\cite{war2000,ban2010}, we introduce the generalized coordinates 
\begin{equation}
q = (z - z_{\mathrm{s}}) / \sigma_{z,0} \quad \mathrm{and} \quad p = (E - E_{\mathrm{s}}) / \sigma_{E, 0} ~.
\label{equ:GeneralizedCoordinates}
\end{equation}
where $z_{\mathrm{s}}$ and $E_{\mathrm{s}}$ denote position and energy of the synchronous particle, $\sigma_{z, 0}$ is the natural bunch length and $\sigma_{E, 0}$ is the natural energy spread. Thereby, the longitudinal charge distribution $\psi$ can be described in the dimensionless phase space spanned by $q$ and $p$, where the origin marks the position of the synchronous particle. The temporal evolution of the normalized distribution $\hat{\psi} (q,p,t)$ is described by the VFP equation
\begin{equation}
\frac{\partial \hat{\psi}}{\partial \theta} + \frac{\partial \mathcal{H}}{\partial p} \frac{\partial \hat{\psi}}{\partial q} - \frac{\partial \mathcal{H}}{\partial q} \frac{\partial \hat{\psi}}{\partial p} = \frac{1}{f_{\mathrm{s},0} \tau_{\mathrm{d}}} \frac{\partial}{\partial p} \left( p \hat{\psi} + \frac{\partial \hat{\psi}}{\partial p} \right) ~,
\label{equ:VFPEquation}
\end{equation}
with time given in multiples of nominal synchrotron periods $\theta = f_{\mathrm{s},0} \, t$, the Hamiltonian $\mathcal{H}$ and the longitudinal damping time $\tau_{\mathrm{d}}$. The inhomogeneous part on the right hand side describes the influence of radiation damping and diffusion. In the absence of collective effects and assuming linear accelerating voltage $V_{\mathrm{RF}}$ and linear momentum compaction factor $\alpha_{\mathrm{c}}$, the Hamiltonian is given as
\begin{equation}
\mathcal{H}_{0}(q,p,t) = \frac{1}{2} \left( q^2 + p^2 \right) ~.
\label{equ:HarmonicHamiltonian}
\end{equation}
The unperturbed system is thus a one-dimensional harmonic oscillator. Collective effects such as CSR self-interaction can be included as a perturbation to the Hamiltonian
\begin{equation}
\mathcal{H}_{\mathrm{c}}(q,p,t) = \int_{q}^{\infty} Q_{\mathrm{c}} \, V_{\mathrm{c}}(q^\prime , t) \mathrm{d}q^\prime ~,
\label{equ:CollectiveHamiltonian}
\end{equation}
where $Q_{\mathrm{c}}$ denotes the charge involved in the perturbation and $V_{\mathrm{c}}(q,t)$ is the potential due to collective effects. In order to calculate the CSR-induced wake potential, it is useful to express the potential in terms of an impedance $Z_{\mathrm{CSR}}(\omega)$
\begin{equation}
V_{\mathrm{CSR}}(q,t) = \int_{-\infty}^{\infty} \widetilde{\rho} (\omega,t) Z_{\mathrm{CSR}}(\omega) e^{i \omega q} \mathrm{d} \omega ~,
\label{equ:WakePotential}
\end{equation}
where $\widetilde{\rho} (\omega)$ denotes the Fourier transformed longitudinal bunch profile.
\begin{figure}[ht]
\centering
\includegraphics{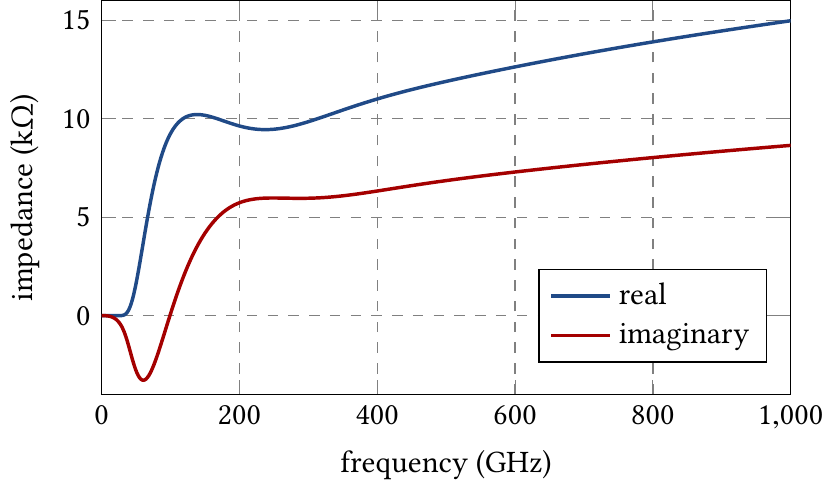}
\caption[CSR parallel plates impedance]
{The process of CSR self-interaction can be described in good approximation by modeling the shielding effect of the vacuum pipe by two parallel plates. Note that the resulting impedance is frequency-dependent and that higher frequencies generally correspond to a larger impedance and thus stronger interaction. Shown is the CSR parallel plates impedance calculated for the KIT storage ring KARA.}
\label{fig:CSRImpedance}
\end{figure}
In general, the exact impedance of a storage ring is not known. However, in the case of CSR-driven dynamics, the approximation of modeling the shielding effect of the beam pipe by two parallel plates~\cite{mur1997} (see FIG.~\ref{fig:CSRImpedance}) has proven to yield results which are quite comparable to experimental data~\cite{bro2016,sch2017,ste2018}. The full Hamiltonian is finally given by
\begin{equation}
\mathcal{H}(q,p,t) = \mathcal{H}_{0}(q,p,t) + \mathcal{H}_{\mathrm{c}}(q,p,t) ~.
\label{equ:FullHamiltonian}
\end{equation}

\begin{figure}[b]
\centering
\includegraphics{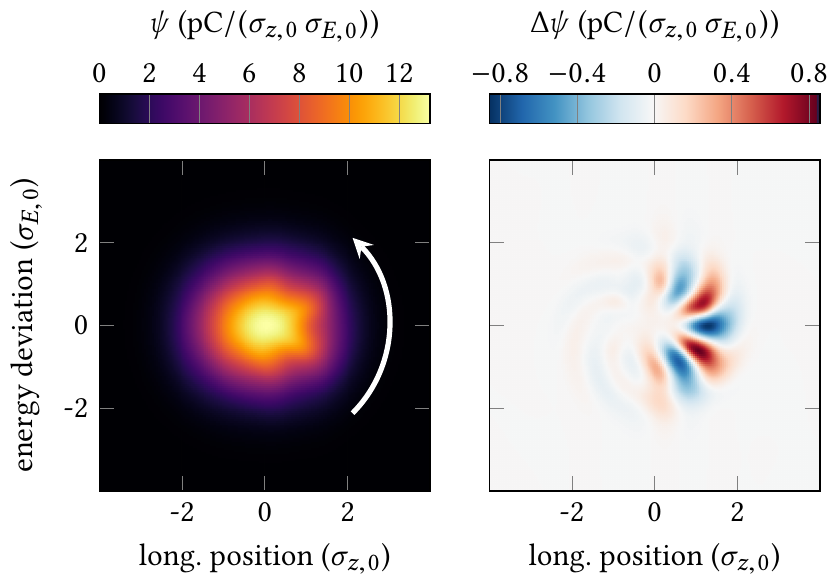}
\caption[Micro-structures in Phase Space]
{Shown is a snapshot of the charge distribution $\psi (q,p,t_{i})$ at time step $t_{i}$ for a simulation of the longitudinal beam dynamics under CSR self-interaction using the VFP solver Inovesa (left). By subtracting the temporal average $\Delta \psi (q,p,t_{i}) = \psi (q,p,t_{i}) - \overline{\psi} (q,p)$ only the non-stationary part of the charge distribution remains, revealing distinct micro-structures (right). Due to synchrotron motion the charge distribution is rotating in phase space and evolving over time.}
\label{fig:PhaseSpaceMicrostructures}
\end{figure}

As of today, there is no analytic solution to the VFP equation for the full Hamiltonian defined in Eq.~(\ref{equ:FullHamiltonian}) and arbitrary parameter settings. Following the approach in~\cite{war2000} however, it can be solved numerically on a discretized grid. To do so, we use the VFP solver Inovesa developed at KIT~\cite{sch2017}. Inovesa is a parallelized open source simulation code based on the mentioned approach, which enables us to do extensive studies using merely standard desktop PCs. The generated simulation data is thoroughly compared to measurements at the KARA storage ring and has shown high qualitative agreement~\cite{ste2018}. As the micro-bunching instability is reproduced with very similar characteristics, the formation and evolution of these micro-structures can be studied in a fully controlled environment facilitating a better understanding of the phenomenon. An exemplary simulated charge distribution $\psi (q,p,t_{i})$ is shown in FIG.~\ref{fig:PhaseSpaceMicrostructures}. By subtracting the temporal average
\begin{equation}
\overline{\psi} (q,p) = \frac{1}{n} \sum_{i}^{n} \psi (q,p,t_{i}) ~,
\label{equ:PhaseSpaceMean}
\end{equation}
the dynamically changing micro-structures become clearly visible. In the following section, the single particle motion corresponding and leading to this charge distribution is examined.

\subsection{\label{subsec:SingleParticleMotion}Single Particle Motion}
\begin{figure}[t]
\centering
\includegraphics{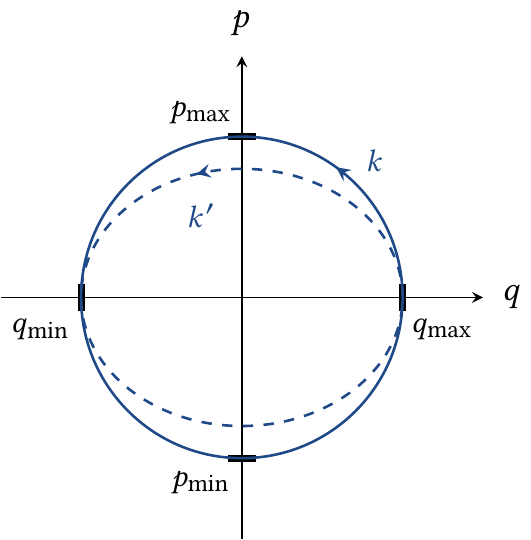}
\caption[1D-harmonic oscillator]
{A small perturbation of the RF slope, i.e.\ the strength of the restoring force $k$ leads to an elliptical particle trajectory in the phase space spanned by the generalized coordinates $q$ and $p$.}
\label{fig:1DHarmonicOscillator}
\end{figure}

In the absence of collective effects the Hamiltonian reduces to Eq.~(\ref{equ:HarmonicHamiltonian}), taking the form of a simple one-dimensional harmonic oscillator. This is because the RF potential acts as a linear restoring force ($\sin (q) \approx q$ for $q$ close to zero)
\begin{equation}
V_{\mathrm{RF}}(q) = - k q \quad ~,
\label{equ:LinearRFPotential}
\end{equation}
with the constant parameter $k$ describing the slope of the RF potential. Neglecting radiation damping and diffusion, this leads to the simple equations of motion
\begin{equation}
\ddot{q} + \frac{k}{\xi} q = 0 \quad , \quad \ddot{p} + \frac{k}{\xi} p = 0 ~,
\label{equ:EquationsOfMotion}
\end{equation}
with the constant scaling parameter $\xi$, and the solution
\begin{align}
q(t) &= a_{0} \cos (\omega t + \varphi_0) \\
\dot{q}(t) &= p(t) = - a_{0} \, \omega \sin (\omega t + \varphi_0) ~,
\label{equ:1dHarmonicSolution}
\end{align}
with $\omega = \sqrt{k / \xi}$, the amplitude $a_{0}$ and the initial phase $\varphi_{0}$. Due to the specific choice of $q$ and $p$ in Eq.~(\ref{equ:GeneralizedCoordinates}), the expression $k / \xi$ simplifies to $\num{1}$, yielding the Hamiltonian in Eq.~(\ref{equ:HarmonicHamiltonian}) and perfectly circular trajectories in phase space. Let us now consider what happens to these trajectories when we introduce a small perturbation to the slope of the RF potential
\begin{equation}
k^\prime = k -\varepsilon \quad \mathrm{with} \quad \varepsilon > 0 ~.
\label{equ:kPerturbation}
\end{equation}
As the restoring force is still linear, the system remains a harmonic oscillator, but now has the altered solution
\begin{align}
q^\prime (t) &= a_{0} \cos (\omega^\prime t + \varphi_0) \\
\dot{q}^\prime (t) &= p^\prime (t) = - a_{0} \, \omega^\prime \sin (\omega^\prime t + \varphi_0) ~,
\label{equ:AlteredHarmonicSolution}
\end{align}
with $\omega^\prime = \sqrt{k^\prime / \xi} = \sqrt{(k - \varepsilon) / \xi}$. While the maximum deviation in $q$ is unaffected
\begin{equation}
\max \, \abs{q^\prime(t)} = \max \, \abs{q(t)} = \abs{a_{0}} ~,
\label{equ:qMaximum}
\end{equation}
the maximum deviation in $p$ is decreased by the perturbation
\begin{equation}
\max \, \abs{p^\prime(t)} = \abs{a_{0} \omega^\prime} < \abs{a_{0} \omega} = \max \, \abs{p(t)} ~.
\label{equ:pMaximum}
\end{equation}
Particle motion in the phase space spanned by the original definitions of $q$ and $p$ in Eq.~(\ref{equ:GeneralizedCoordinates}) is thus elliptical, as illustrated in FIG.~\ref{fig:1DHarmonicOscillator}, and of altered periodicity 
\begin{equation}
\abs{\omega^\prime} < \abs{\omega} ~.
\label{equ:AlteredPeriodicity}
\end{equation}

In order to examine the perturbation of this simple harmonic system by CSR self-interaction, we need to consider the additional potential introduced in Eq.~(\ref{equ:WakePotential}).

\section{\label{sec:ParticleMotionBelowThreshold}Particle Motion Below Threshold}
\begin{figure}[ht]
\centering
\includegraphics{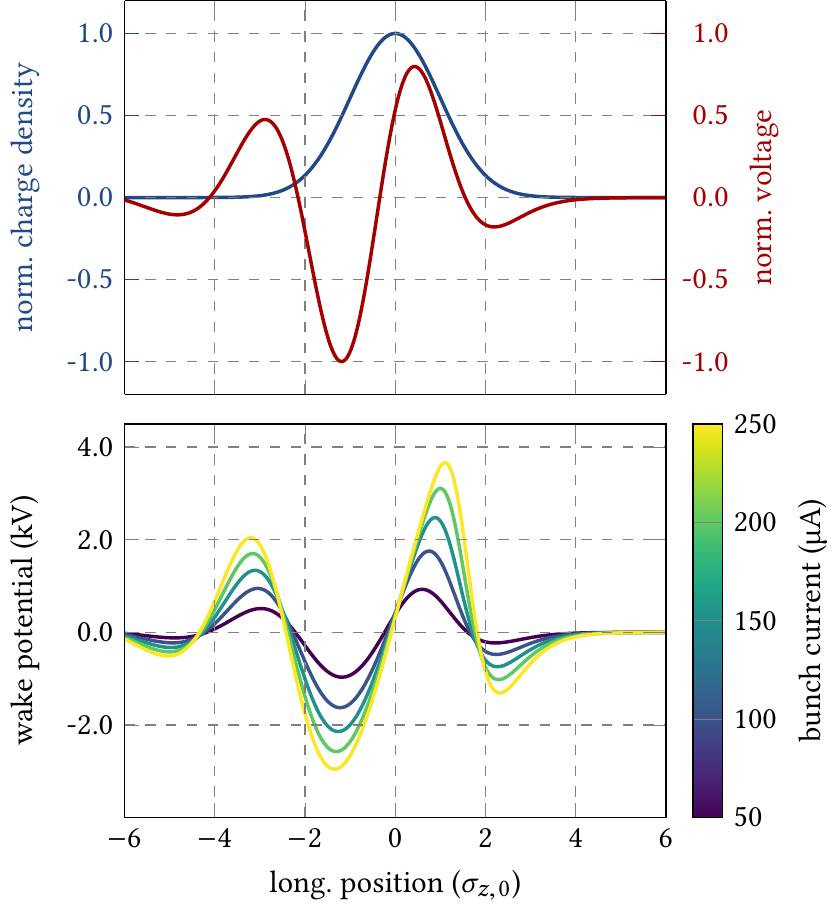}
\caption[Wake potential]
{CSR wake potential for a Gaussian bunch profile (red and blue, top) and for the bunch currents $I = (\num{50},\num{100},\num{150},\num{200},\num{250}) \, \si{\micro\ampere}$ below the instability threshold of $I_{\mathrm{th}} = \SI{260}{\micro\ampere}$ (bottom). Shown are the respective temporal averages for simulations of the KIT storage ring KARA using Inovesa (including damping and diffusion).}
\label{fig:WakePotential}
\end{figure}

Given the parallel plates impedance $Z_{\mathrm{CSR}}$ shown in FIG.~\ref{fig:CSRImpedance}, the wake potential of a Gaussian bunch profile takes the form depicted in the upper part of FIG.~\ref{fig:WakePotential}. While such a perfectly Gaussian electron distribution only exists in the zero current limit, a higher bunch current leads to an increased perturbation strength and thus distortion of the Gaussian shape.
\begin{figure}[ht]
\centering
\includegraphics{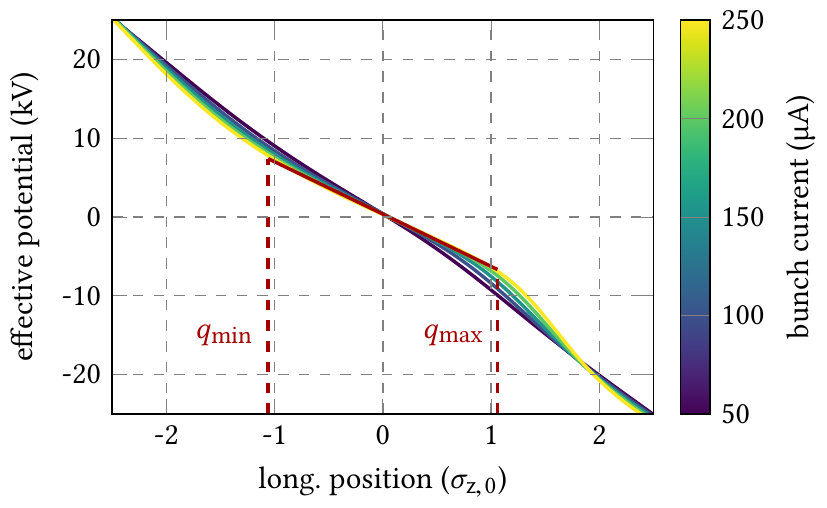}
\caption[Effective potential below threshold]
{Combining the RF potential $V_{\mathrm{RF}}(q)$ and the CSR wake potential $V_{\mathrm{CSR}}(q)$ yields an effective potential $V_{\mathrm{eff}}(q)$ that the electrons are exposed to during their revolution in the storage ring. Close to the synchronous position $q=0$, the potential can be approximated by a linear function (shown as solid red line for an exemplary particle with an amplitude of roughly $q_{\mathrm{max}} = \num{1.1}$). For larger deviations from $q=0$ the linear approximation becomes less accurate, but still provides a useful estimate of the perturbed potential.}
\label{fig:EffectivePotentialBelowThreshold}
\end{figure}
Yet, below the threshold current, the distribution still remains fairly stationary $\psi (q,p,t) \approx \psi (q,p)$, which corresponds to a stationary wake potential as shown in the lower part of FIG.~\ref{fig:WakePotential} for a range of different bunch currents. It should be noted that the general shape of the wake potential is still similar to that of a Gaussian shaped bunch up until right below the threshold current of $I_{\mathrm{th}} = \SI{260}{\micro\ampere}$, where the wake potential is no longer stationary. In order to investigate the effect of this additional potential on the single particle motion, we introduce the effective potential
\begin{equation}
V_{\mathrm{eff}}(q) = V_{\mathrm{RF}}(q) + V_{\mathrm{CSR}}(q) ~,
\label{equ:EffectivePotential}
\end{equation}
combining the linear RF potential $V_{\mathrm{RF}}(q)$ and the CSR wake potential $V_{\mathrm{CSR}}(q)$. Thereby, we average over the full storage ring and neglect the fact that these two types of interaction with external fields are happening at different positions in the storage ring (e.g.\ RF cavities and bending magnets). This simplification is reasonable as the synchrotron motion happens at a much slower time scale than a single passage of the storage ring ($f_{\mathrm{rev}} \gg f_{\mathrm{s},0}$).

A single particle moving in phase space is now subject to the effective potential $V_{\mathrm{eff}}(q)$ over the interval $\left[ q_{\mathrm{min}}, q_{\mathrm{max}} \right]$, where $q_{\mathrm{min}}$ and $q_{\mathrm{max}}$ denote the maximum deviations from the longitudinal position of the synchronous particle (see FIG.~\ref{fig:1DHarmonicOscillator}). By approximating $V_{\mathrm{eff}}(q)$ as a linear function on the given interval
\begin{equation}
V_{\mathrm{eff}}(q) \approx - k^{\prime} \, q ~, \quad q \in \left[ q_{\mathrm{min}}, q_{\mathrm{max}} \right] ~,
\label{equ:LinearApproximatedPotential}
\end{equation}
as illustrated in FIG.~\ref{fig:EffectivePotentialBelowThreshold}, the single particle motion is still harmonic below the threshold current, with the strength of the restoring force $k^{\prime}$ being dependent on $q_{\mathrm{min}}$ and $q_{\mathrm{max}}$. While the linear approximation seems quite suitable for particles close to $q=0$, the approximation is getting inaccurate for oscillation amplitudes larger than $\pm \sigma_{z, 0}$. However, we're mostly interested in describing the potential near $q=0$ where the majority of the charge is located, and we will see that this simple model yields a reasonable approximation nonetheless. As is apparent from FIG.~\ref{fig:EffectivePotentialBelowThreshold}, the CSR wake potential acts as a perturbation of the RF slope with the strength of the perturbation being dependent on the amplitude of the particle's oscillation. According to Eq.~(\ref{equ:kPerturbation}-\ref{equ:AlteredPeriodicity}), this results in a position-dependent ellipticity of particle trajectories in phase space. Analogously, the oscillation frequency varies as a function of the particle's position. 

\begin{figure}[b]
\centering
\includegraphics{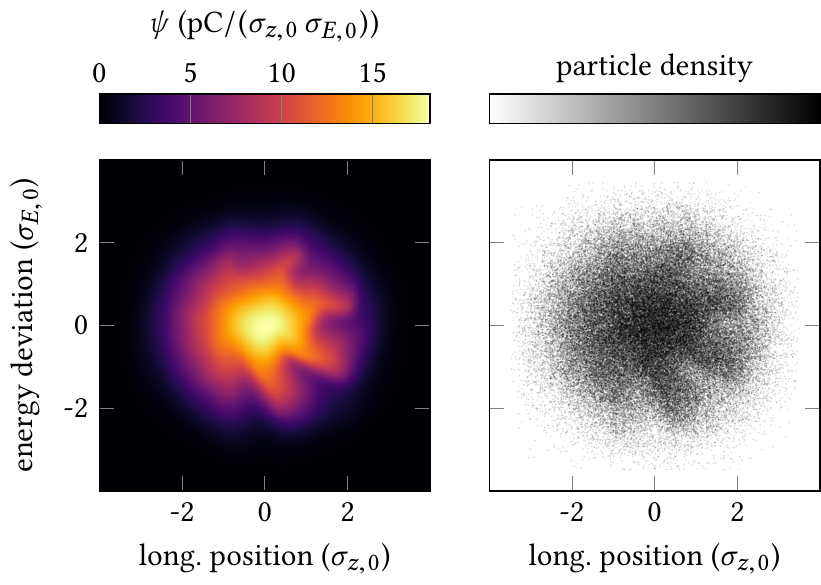}
\caption[Particle Distribution]
{In order to examine single particle trajectories, the initial charge density $\psi (q,p,t_{0})$ (left) is modeled by a distribution of $n = \num{e5}$ macro-particles (right). The single particle motion is then simulated using the passive particle tracking implemented in Inovesa.}
\label{fig:ModelingParticleDistribution}
\end{figure}

\begin{figure}[b]
\centering
\includegraphics{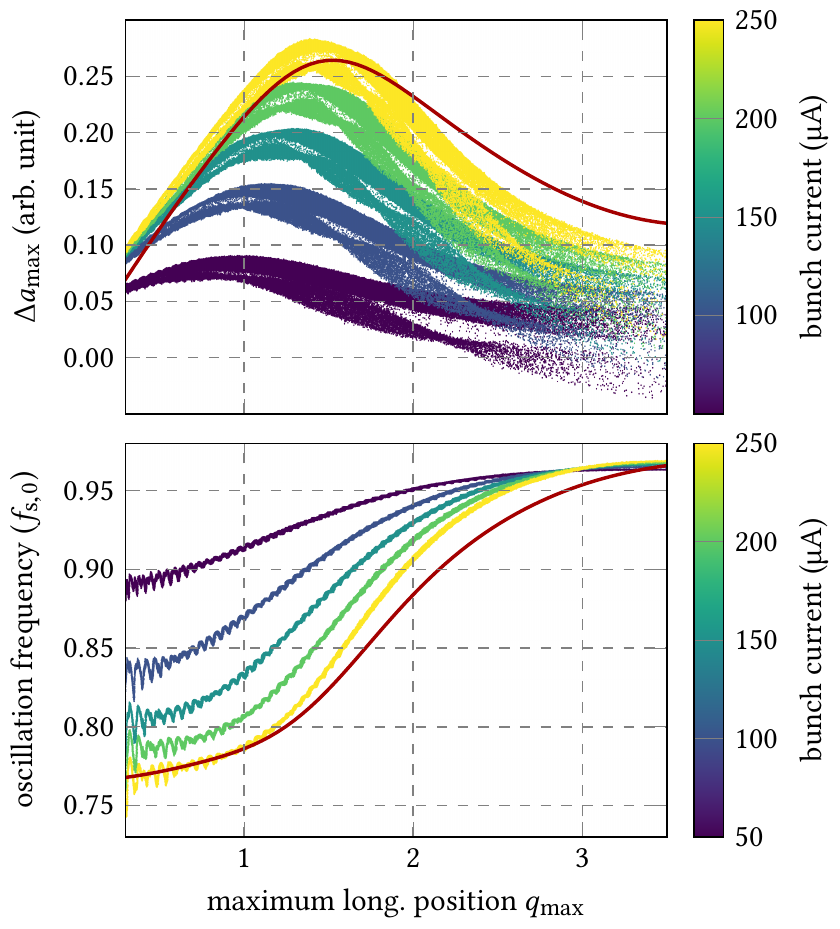}
\caption[Particle Trajectory Statistics]
{Amplitude differences (top) and oscillation frequencies (bottom) of $n = \num{e5}$ particle trajectories simulated with Inovesa for several bunch currents below the instability threshold. Note that the oscillation frequencies scatter significantly less than the amplitude differences. The small oscillations of the frequencies in the range of $\num{0} < q_{\mathrm{max}} < \num{1}$ are presumed numerical artifacts of the simulation and data analysis. The solid red lines depict predictions based on the linear approximation of $V_{\mathrm{eff}}(q)$ for the bunch current $I=\SI{250}{\micro\ampere}$.}
\label{fig:ParticleTrajectoryStatisticsBelowThreshold}
\end{figure}

We can verify these insights using the passive particle tracking method\footnote{The computation of the CSR wake potential is still based on the charge distribution function $\psi (q,p,t)$, particles are tracked accordingly.} that was recently added to Inovesa~\cite{sch2018}. To that end, the initial charge distribution $\psi (q,p,t_{0})$ is modeled by a particle ensemble of $n = \num{e5}$ macro-particles (see FIG.~\ref{fig:ModelingParticleDistribution}).
Subsequently, the temporal evolution under the influence of the CSR wake potential is calculated simultaneously for both, the charge and the particle distribution. The thus obtained simulation results for the particle distribution are displayed in FIG.~\ref{fig:ParticleTrajectoryStatisticsBelowThreshold}. Here, the upper part shows the difference of the maximum amplitude in $q$ and $p$
\begin{equation}
\Delta a_{\mathrm{max}}(q_{\mathrm{max}}) = q_{\mathrm{max}} - p_{\mathrm{max}} ~,
\label{equ:AmpitudeDifference}
\end{equation}
for the individual particles as a function of their maximum longitudinal deviation $q_{\mathrm{max}}$. The particle trajectories clearly deviate from $\Delta a_{\mathrm{max}} = 0$, which would correspond to a circular trajectory, and thereby show the expected position-dependent ellipticity. While the trajectories are already elliptical close to the origin, the maximum difference in amplitude is reached at values of $q_{\mathrm{max}}$ in the range of $\num{1.0}$ to $\num{1.5}$ depending on the bunch current. Both the maximum value of $\Delta a_{\mathrm{max}}$ and the corresponding longitudinal position are increasing with higher bunch currents due to the increased strength of the perturbation. For particles with larger deviation from the synchronous particle the amplitude difference reduces again, indicating a trend to more circular shaped trajectories for larger amplitudes.
Additionally, the lower part of FIG.~\ref{fig:ParticleTrajectoryStatisticsBelowThreshold} displays the corresponding oscillation frequencies. As expected from Eq.~(\ref{equ:kPerturbation}-\ref{equ:AlteredPeriodicity}), the individual synchrotron frequencies of particles close to $q=0$ are significantly lower than the nominal synchrotron frequency $f_{\mathrm{s},0}$ due to the perturbation of the linear restoring force. Yet, this difference diminishes for particle trajectories with larger amplitude yielding different oscillation frequencies dependent on the position of the particles within the bunch. Finally, we can directly compare these results with predictions based on the linear approximation of $V_{\mathrm{eff}}(q)$. Therefore, we use the estimated value of $k^\prime (q_{\mathrm{max}})$ to determine the oscillation frequency at that position
\begin{equation}
f_{\mathrm{s}}(q_{\mathrm{max}}) = \omega_{\mathrm{s}}(q_{\mathrm{max}}) / 2 \pi \approx \sqrt{k^\prime (q_{\mathrm{max}})} / \zeta ~,
\label{equ:ParticleFrequencyPrediction}
\end{equation}
where $\zeta$ is just used for normalization purposes. Given an estimate of the oscillation frequency and using the approximation $p_{\mathrm{max}} \approx q_{\mathrm{max}} \omega_{\mathrm{s}}(q_{\mathrm{max}})$, we can easily calculate the expected difference in amplitude
\begin{equation}
\Delta a_{\mathrm{max}}(q_{\mathrm{max}}) \approx q_{\mathrm{max}} \left[ 1 - \omega_{\mathrm{s}}(q_{\mathrm{max}}) \right] ~.
\label{equ:AmplitudeDifferencePrediction}
\end{equation}
The thus calculated estimates are shown as solid red lines in FIG.~\ref{fig:ParticleTrajectoryStatisticsBelowThreshold} for the case of $I=\SI{250}{\micro\ampere}$. Clearly, the linear approximation of the effective potential in Eq.~(\ref{equ:LinearApproximatedPotential}) is already sufficient for describing a major part of the perturbation by the CSR wake potential and its effect on the synchrotron motion of single particles. As expected, the estimates of $f_{\mathrm{s}}(q_{\mathrm{max}})$ and $\Delta a_{\mathrm{max}}(q_{\mathrm{max}})$ deviate from the simulated trajectories for larger values of $q_{\mathrm{max}}$ due to the inaccuracy of the linear approximation of $V_{\mathrm{eff}}(q)$ for values further away from $q=0$. The general shape however, can still be reproduced.

\begin{figure}[b]
\centering
\includegraphics{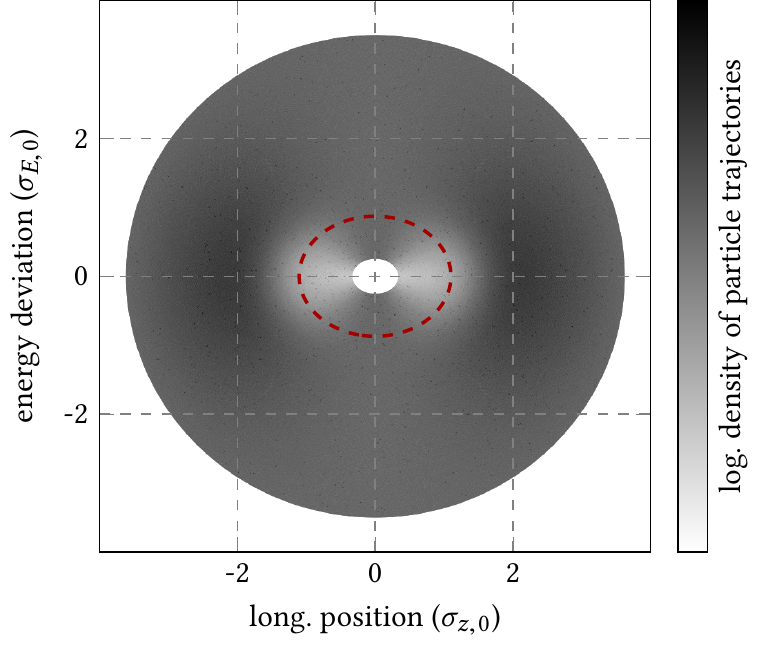}
\caption[Toy study of elliptical trajectories]
{Visualization of the effect of the position-dependent elliptical trajectories in phase space on the resulting charge density. Shown are perfectly elliptical trajectories of $\num{e3}$ particles with uniformly distributed energies and an amplitude difference given by the estimate shown as solid red line in FIG.~\ref{fig:ParticleTrajectoryStatisticsBelowThreshold}. The large number of trajectories helps to visualize the varying density of trajectories. In addition, the dashed red line depicts a single particle trajectory with $q_{\mathrm{max}} \approx 1.1$. }
\label{fig:ToyStudyEllipses}
\end{figure}

In order to understand the implications of these modified single particle trajectories, let us consider an ensemble of particles with uniformly distributed energies and perfectly elliptical trajectories. Furthermore, let the ellipticity be determined according to the position-dependent amplitude difference $\Delta a_{\mathrm{max}} (q_{\mathrm{max}})$ shown as solid red line in FIG.~\ref{fig:ParticleTrajectoryStatisticsBelowThreshold}. As is apparent from the visualization in FIG.~\ref{fig:ToyStudyEllipses}, this leads to a non-uniform distribution of particle trajectories in phase space. In particular, two distinguished locations of lower particle concentration are visible close to the origin. Similarly, though harder to identify by eye, there are two locations of higher particle concentration at larger oscillation amplitudes (roughly at $q \approx \pm \num{2}$). This general pattern is a direct consequence of the basic shape of the CSR wake potential shown in FIG.~\ref{fig:WakePotential}. The CSR-induced perturbation of the RF potential thus breaks the homogeneity in phase space and creates local particle densities that form a quadrupole-like modulation of the longitudinal charge distribution. This inhomogeneity introduces a higher frequency component to the longitudinal bunch profile and may thereby initially seed the formation of micro-structures, i.e.\ kick off the micro-bunching instability. Note that the general notion of dense particle trajectories leading up to a distinct charge modulation within the bunch resembles the caustic expression adopted for micro-bunching phenomena in linear accelerators~\cite{cha2016}.

\begin{figure}[b]
\centering
\includegraphics{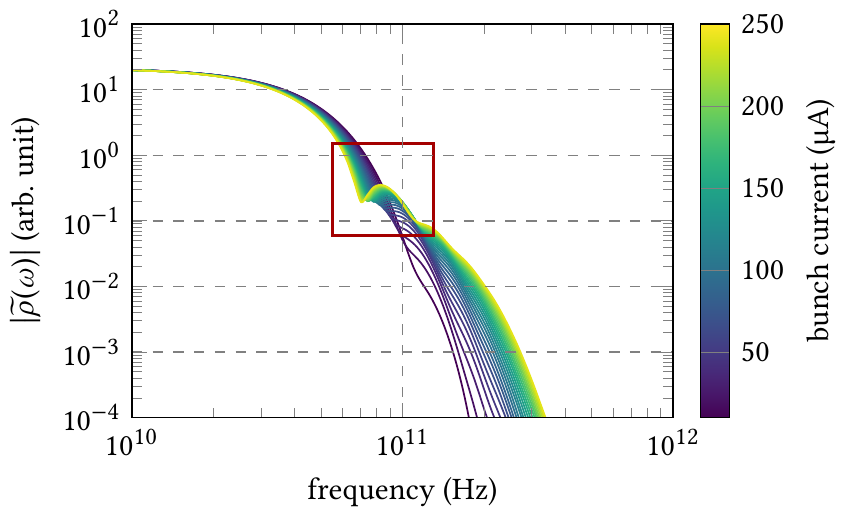}
\caption[Quadrupole-mode in Fourier transformed bunch profiles]
{Shown is the magnitude of the Fourier transformed bunch profile $\abs{\widetilde{\rho}(\omega)}$ for a range of bunch currents below the instability threshold of $I_{\mathrm{th}}=\SI{260}{\micro\ampere}$. The red rectangle marks the additional frequency component (at roughly $\SI{85}{\giga\hertz}$) that is excited due to the perturbation by the CSR wake potential.}
\label{fig:FFTProfiles}
\end{figure}

We can further verify this initial excitation of a quadrupole-like mode by examining the Fourier transformed longitudinal bunch profiles across different bunch currents below the instability threshold. As is apparent from FIG.~\ref{fig:FFTProfiles}, the current-dependent perturbation by the CSR wake potential introduces a higher frequency component to the longitudinal charge distribution. For the used parameter settings this frequency is found at about $\SI{85}{\giga\hertz}$ corresponding to a modulation of the bunch profile at the wavelength
\begin{equation}
\lambda \approx c / \SI{85}{\giga\hertz} \approx \num{2.2} \, \sigma_{z,0} ~,
\label{equ:WavelengthQuadrupoleMode}
\end{equation}
which is roughly the distance of the two expected distinguished locations of decreased charge density in phase space. Note that the micro-structures occurring above the threshold current correspond to a significantly higher frequency (here, roughly $\SI{150}{\giga\hertz}$), so the peak in FIG.~\ref{fig:FFTProfiles} can clearly be attributed to the initial quadrupole mode.

Finally, we would like to conclude this section by pointing out that the additional weak instability~\cite{ban2010,ban2005,bro2019}, that occurs for specific parameter settings of the storage ring, is usually observed with an instability frequency of $f_{\mathrm{inst}} \approx 2 \, f_{\mathrm{s},0}$. This corresponds to a non-stationary quadrupole mode on the longitudinal charge distribution, which aligns remarkably well with the stationary deformation below the threshold current discussed here. Moreover, measurements of such a quadrupole-like deformation of the longitudinal charge distribution were already reported in~\cite{pod1997}, albeit for bunch currents above the instability threshold.

\section{\label{sec:ParticleMotionAboveThreshold}Particle Motion above Threshold}
In the previous section we discussed the perturbation of single particle synchrotron motion by the stationary CSR wake potential below the threshold current. Here, the linear approximation of the effective potential in Eq.~(\ref{equ:LinearApproximatedPotential}) yields a simple model which sufficiently describes major aspects of the resulting particle trajectories and thereby facilitates understanding the implications of this perturbation. Essentially, the underlying longitudinal dynamics can be considered a simple one-dimensional harmonic oscillator with a position-dependent perturbation of the linear restoring force.
\begin{figure}[b]
\centering
\includegraphics{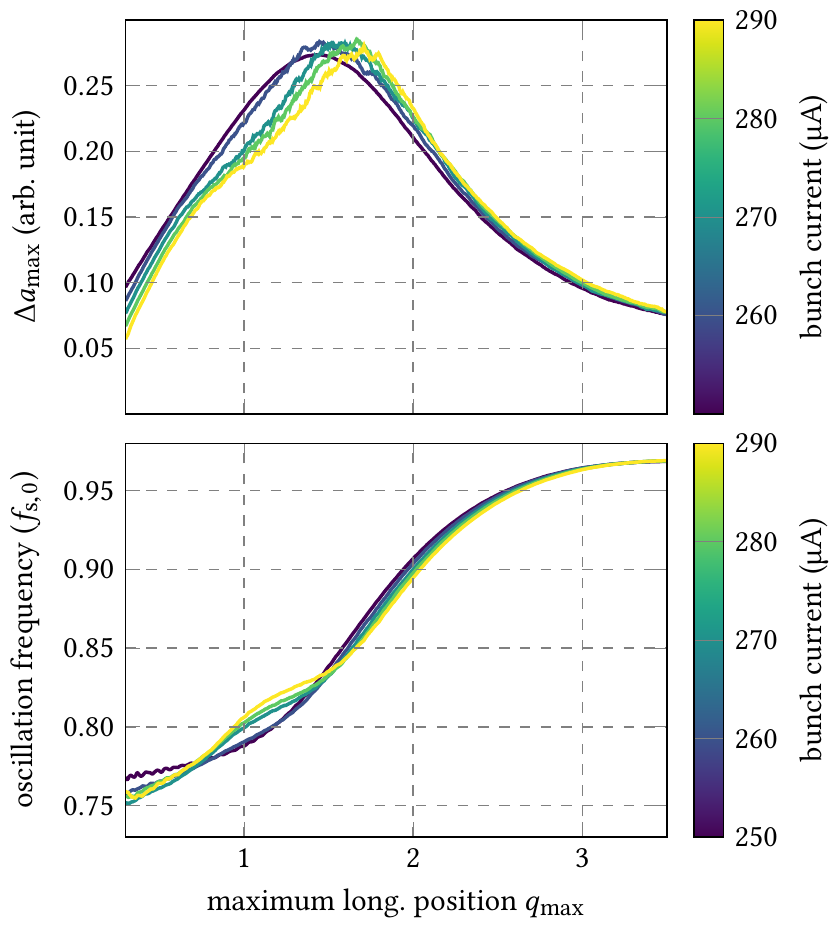}
\caption[Particle Trajectory Statistics above Threshold]
{Shown are the amplitude difference (top) and oscillation frequency (bottom) of particles trajectories for the bunch currents $I = (\num{250},\num{260},\num{270},\num{280},\num{290}) \, \si{\micro\ampere}$. As the data scatters a lot, only a moving average over $q_{\mathrm{max}}$ is displayed to enable a comparison between different currents.}
\label{fig:ParticleTrajectoryStatisticsAboveThreshold}
\end{figure}
\begin{figure*}[ht]
\centering
\includegraphics{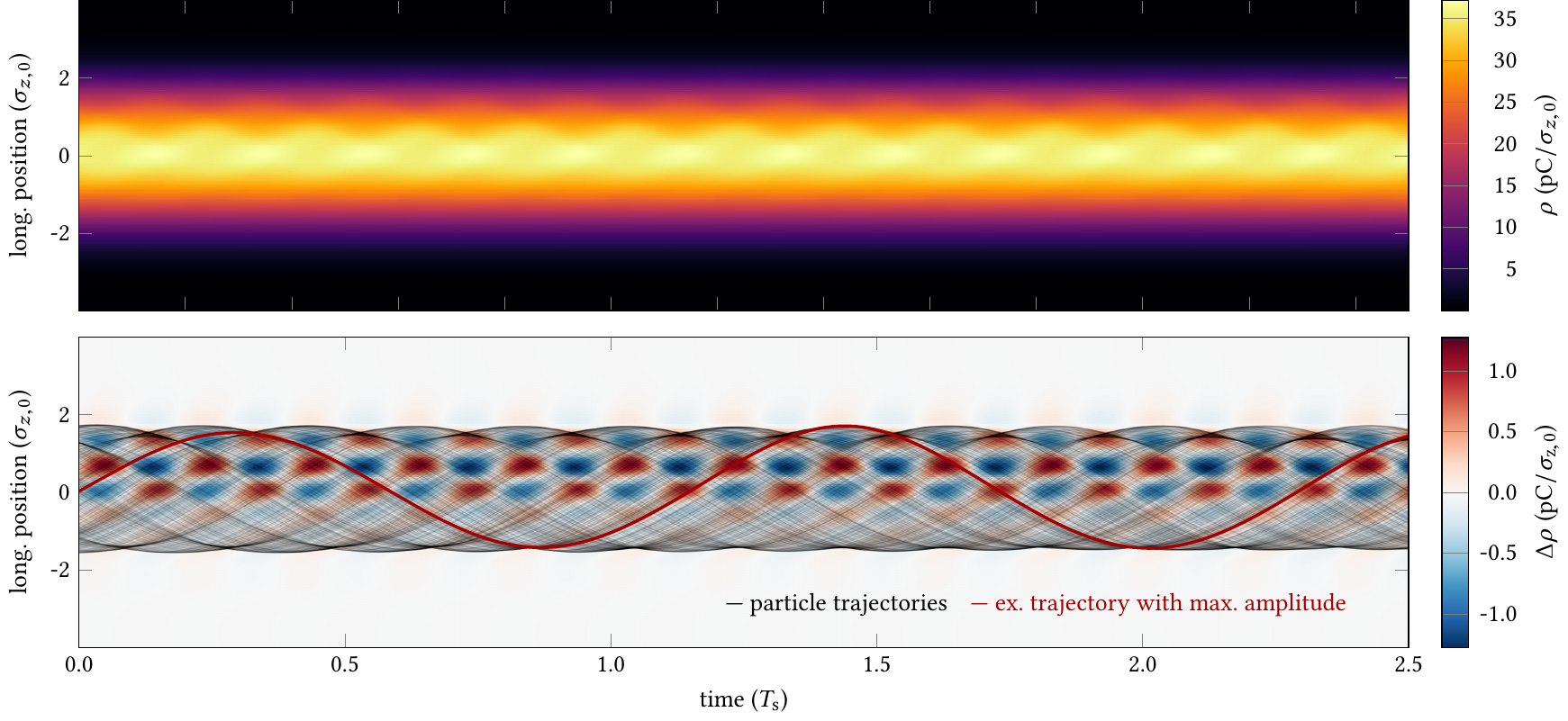}
\caption[Particle Trajectories and Evolution of Bunch Profiles]
{Temporal evolution of the longitudinal bunch profile (top) and its difference to the temporal average $\Delta \rho (q,t) = \rho (q,t) - \overline{\rho}(q)$ (bottom) for the bunch current $I=\SI{290}{\micro\ampere}$. In order to visualize the corresponding single particle motion $\num{e3}$ particle trajectories are plotted on top with an opacity of $\num{0.05}$, displaying a distinct modulation of the maximum deviation in the longitudinal position. The maximum oscillation amplitude is reached for particles that are exposed to the additional wake potential caused by the local charge modulation (solid red curve).}
\label{fig:ParticleTrajectoriesAndBunchProfiles}
\end{figure*}
Above the instability threshold, the longitudinal charge distribution as well as the CSR wake potential are not stationary anymore, which makes this simple model no longer applicable. Nevertheless, we will see how the notion of a perturbed restoring force extends to the dynamics just above the instability threshold and eventually motivates an approach to potential control of the micro-bunching instability. The provided analysis builds upon extensive, prior studies of the micro-bunching dynamics above the instability threshold at KARA. It extends the gained understanding of the underlying longitudinal beam dynamics and offers a new perspective on the interpretation of previous observations.

In order to examine the single particle trajectories above the instability threshold, the initial charge density $\psi (q,p,t_{0})$ is again modeled by a distribution of macro-particles and passively tracked using Inovesa. The upper part of FIG.~\ref{fig:ParticleTrajectoryStatisticsAboveThreshold} shows the amplitude difference $\Delta a_{\mathrm{max}}(q_{\mathrm{max}})$ of the resulting particle trajectories for a range of bunch currents in comparison to a current below the threshold ($I=\SI{250}{\micro\ampere}$, violet line). Due to the more complex dynamics and the variation of the individual particle trajectories in time, the data scatters a lot. In order to enable a comparison between multiple currents nonetheless, a moving average over $q_{\mathrm{max}}$ is displayed. The elliptical shape is still apparent and seems quite comparable to the results below the threshold current in FIG.~\ref{fig:ParticleTrajectoryStatisticsBelowThreshold}. The position of the maximum amplitude difference $\Delta a_{\mathrm{max}}(q_{\mathrm{max}})$ however, is slightly shifting to larger values. This implies that the local charge accumulation (see FIG.~\ref{fig:ToyStudyEllipses}) also shifts to a position further away from the origin. Particularly interesting is the change in the distribution of the corresponding oscillation frequencies that is displayed below. While the higher bunch current leads to an abrupt change of the oscillation frequencies in the interval $q_{\mathrm{max}} \in \left[ \num{0.3},\num{1.5} \right]$, it has a much smaller effect on the oscillation frequencies of particles with larger amplitudes. This region of $q_{\mathrm{max}} \in \left[ \num{0.3},\num{1.5} \right]$ is precisely where the micro-structures occur in phase space and hints to the additional wake potential caused by the corresponding charge modulation. To factor in the temporal dynamics above the instability threshold, we need to find yet a different form of visualizing the individual particle trajectories. Following the example of prior work, e.g.~\cite{rot2016,fun2019,bro2019syn}, the upper part of FIG.~\ref{fig:ParticleTrajectoriesAndBunchProfiles} thus displays the temporal evolution of the longitudinal bunch profile over a time period of two and a half synchrotron periods for the bunch current $I=\SI{290}{\micro\ampere}$. By close examination, the periodic modulation of the charge density due to the occurring micro-structures can already be identified. Nevertheless, in analogy to FIG.~\ref{fig:PhaseSpaceMicrostructures}, the lower part shows again the difference to the temporal average $\overline{\rho} (q)$, which displays the micro-bunching dynamics much more explicitly. In order to examine how the motion of single particles relates to these micro-bunching dynamics $\num{e3}$ particle trajectories are plotted on top with an opacity of $\num{0.05}$. These trajectories are deliberately chosen to have an average radius in phase space
\begin{equation}
\overline{r} = \frac{1}{n} \sum_{i=1}^{n} r_{t_{i}} = \frac{1}{n} \sum_{i=1}^{n} \sqrt{q_{t_{i}}^2 + p_{t_{i}}^2} ~,
\label{equ:AveragePhaseSpaceRadius}
\end{equation}
which is comparable to the estimated distance of the micro-structures from the origin.

In the following subsections, we will examine different aspects of particle motion above the instability threshold using the insights of previous sections.

\subsection{\label{subsec:HeadTailAsymmetry}Head-Tail Asymmetry}
Particularly intriguing is the distinct sinusoidal modulation of the maximum amplitude $q_{\mathrm{max}}$ that is visible at the head of the bunch ($q>0$). This modulation is perfectly synchronized to the micro-structure dynamics in the charge density and reaches its extrema at exactly the same positions in time. However, similarly to the charge modulation, it predominantly occurs at the head and diminishes towards the tail of the bunch. The maximum amplitude in $q$ is reached when a particle travels on a trajectory that leads to its exposure to an additional contribution in the CSR wake potential caused by the local charge modulation. Particles traveling exactly along the position of the maximum local charge density (red areas) while passing through $q \in \left[ 0, 2 \right]$ are subsequently driven to the largest deviation in $q$ (illustrated by the solid red curve in FIG.~\ref{fig:ParticleTrajectoriesAndBunchProfiles}). Similarly, particles passing through the minima (blue areas) end up closest to the origin.

This asymmetry can be explained by the different effects a local structure at different positions in the charge density has on the restoring force. For $q>0$, a positive contribution to the effective potential $V_{\mathrm{eff}}$ results in a further decrease of the restoring force and thus drives particles further outside in phase space. This amplifies the inhomogeneity and can thereby drive and support the local charge modulation. In contrast, a positive contribution at $q<0$ partially recovers the strength of the restoring force and focuses the particles towards the center of the charge density, reducing the inhomogeneity and damping the local structure.

\subsection{\label{subsec:FormationOfStructures}Formation of Micro-Structures}
The previous subsection illustrated why the micro-structures are more pronounced at the head of the bunch rather than the tail. Simultaneously, it explains how single particle motion is leading up to these local charge modulations. Particles are driven outside in phase space, cause an excess of charge at that position and thereby form the occurring micro-structures.

\begin{figure}[t]
\centering
\includegraphics{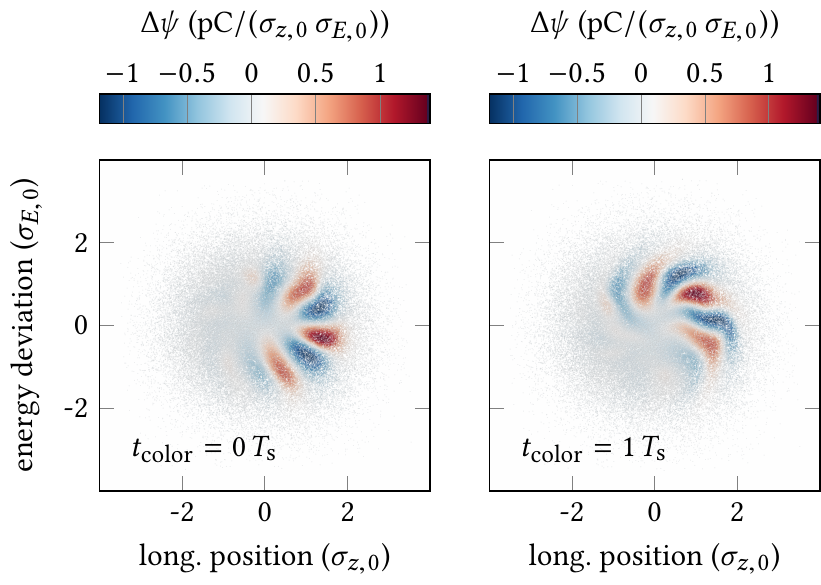}
\caption[Formation of Micro-Structures]
{Shown is the particle distribution at time step $t=\num{0} \, T_{\mathrm{s}}$ for the bunch current $I=\SI{290}{\micro\ampere}$. On the left hand side, each individual particle is colored according to the corresponding relative charge density $\Delta \psi (q,p,t_{\mathrm{color}}=\num{0} \, T_{\mathrm{s}})$ at that particle's location (opacity of $0.15$). On the right hand side, the color assignment is adjusted to match $\Delta \psi (q,p,t_{\mathrm{color}}=\num{1} \, T_{\mathrm{s}})$ instead. It thereby illustrates which particles will form the micro-structures one synchrotron period after this time step.}
\label{fig:ColoredParticleDistribution}
\end{figure}

We would now like to further investigate how the motion of individual particles relates to the motion of the observed micro-structures. Therefore, we'll take a look at the location of particles one synchrotron period before they form a local structure. To do so, FIG.~\ref{fig:ColoredParticleDistribution} displays the particle distribution at time step $t=\num{0}\,T_{\mathrm{s}}$ in two different ways. On the left hand side each individual particle is colored according to the relative charge density at this time step
\begin{equation}
\mathrm{Color}(n=i) \leftarrow \Delta \psi (q_{n=i},p_{n=i},t_{\mathrm{color}}=\num{0} \, T_{\mathrm{s}}) ~,
\label{equ:ParticleColorAssignment}
\end{equation}
where $n=i$ is the particle index and $(q_{n=i},p_{n=i})$ denotes its location in phase space at time $t_{\mathrm{color}}$. The color assignment on the right hand side is adjusted to match the relative charge density one synchrotron period afterwards $\Delta \psi (q_{n=i},p_{n=i},t_{\mathrm{color}}=\num{1} \, T_{\mathrm{s}})$ instead. The distribution on the right hand side thus shows where the particles forming the micro-structures at time step $t=\num{1}\,T_{\mathrm{s}}$ were one synchrotron period before. Thereby, we are able to analyze where the particles forming the local structures come from and whether or not they stay within these structures. Clearly, the two distributions look quite different. This implies that the particles forming the structures at $t=\num{0}\,T_{\mathrm{s}}$ do not necessarily participate in forming the same kind of structure one synchrotron period later. They might e.g.\ travel from the position of a local maximum (red) to position of a local minimum (blue) or vice versa. This effect was already observed in previous studies~\cite{sch2018diss}.

FIG.~\ref{fig:ColoredParticleDistribution} illustrates conceptually that the formation of these micro-structures is not merely caused by the resonant motion of single particles, but rather by the collective effect of many particles traveling on varying trajectories. Moreover, during the analysis presented here, we found that the relation between the motion of individual particles and the occurring micro-structures can vary significantly across different bunch currents and parameter settings.

\subsection{\label{subsec:InstabilityFrequency}Instability Frequency}
As mentioned in the introduction, the occurrence of micro-structures in the longitudinal phase space results in fluctuations of the emitted CSR power. Right above the instability threshold, this fluctuation is typically dominated by one characteristic frequency, which we will denote with $f_{\mathrm{inst}}$. As reported by different facilities, e.g.~\cite{ban2005,bil2016,bro2020diss}, this frequency is usually observed close to an integer multiple of the nominal synchrotron frequency 
\begin{equation}
f_{\mathrm{inst}} \approx m \, f_{\mathrm{s},0} ~,
\label{equ:InstabilityFrequencyEstimate}
\end{equation}
but may deviate significantly depending on the parameter settings~\cite{bil2016,bro2020diss}. In this subsection, we aim to illustrate how this frequency originates from the micro-bunching dynamics in phase space.

The integrated power of the emitted CSR $P_{\mathrm{CSR}}(t)$ is solely determined by the projection of the charge density $\psi(q,p,t)$ on the longitudinal axis, i.e.\ the longitudinal bunch profile $\rho(q,t)$ and the CSR impedance. Different charge distributions in phase space at different time steps $t_{i,j}$ thus result in the exact same value of the radiated power if they correspond to identical longitudinal profiles $\rho(q,t_{i}) \equiv \rho(q,t_{j})$. Consequently, the periodic structure of the fluctuating CSR power can be explained by a repetitive sequence of the longitudinal profiles. The most trivial way to produce similar longitudinal profiles at different time steps is by having the charge densities in phase space be similar as well
\begin{equation}
\psi(q,p,t_{i}) \simeq \psi(q,p,t_{j}) ~.
\label{equ:SimilarChargeDensities}
\end{equation}
In the simulations with Inovesa we find this to always be the case. The charge densities separated by one period of the instability frequency $\Delta t = 1/f_{\mathrm{inst}}$ are nearly indistinguishable by eye and only differ by small numeric values. The observed instability frequency is thus directly determined by the periodic behavior of the micro-bunching dynamics and the corresponding time interval. However, as established above, the propagation of these micro-structures in phase space and in time is a non-trivial subject. With the varying oscillation frequencies of the single particle trajectories (e.g.\ illustrated in FIG.~\ref{fig:ParticleTrajectoryStatisticsAboveThreshold}) and the collective formation of the occurring micro-structures, very little can be deduced about the oscillation frequency of the micro-structures themselves. Empirically, we found that this can deviate up to $\num{20} \, \%$ from the nominal synchrotron frequency. In that case, the simple estimate in Eq.~(\ref{equ:InstabilityFrequencyEstimate}) is no longer applicable. In particular, trying to estimate the integer $m$ via
\begin{equation}
m \approx f_{\mathrm{inst}} / f_{\mathrm{s},0}
\label{equ:ModeEstimate}
\end{equation}
will yield inconsistent and unreliable results as the micro-structures may propagate at differing frequencies.

Lastly, note that the interpretation of $m$ as a simple azimuthal mode number is difficult to align with the head-tail asymmetry of the CSR self-interaction discussed in subsection~\ref{subsec:HeadTailAsymmetry}. This asymmetry explains why the micro-structures always form at the head of the bunch and are more pronounced there. The instability frequency observed in the emitted CSR power is simply determined by the repetition rate of this formation process.

\begin{figure}[t]
\centering
\includegraphics{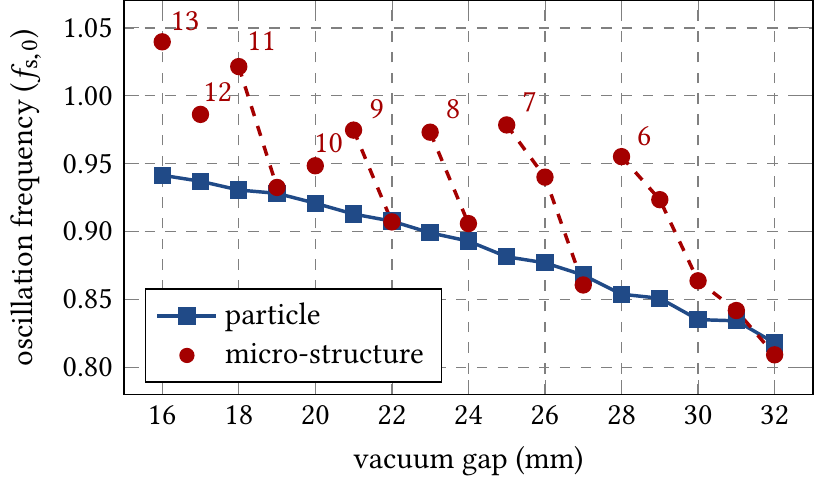}
\caption[Dependence on Shielding]
{Modifying the CSR shielding by varying the vacuum gap leads to altered micro-bunching dynamics in the longitudinal phase space. While the single particle frequency at the micro-structure position (blue squares) increases smoothly with reduced vacuum gap, the oscillation frequency of the micro-structures themselves (red circles) grows much faster and shows an abrupt change when an additional structure is observed.}
\label{fig:DependenceOnShielding}
\end{figure}

\subsection{\label{subsec:DependenceOnShielding}Dependence on Shielding}
Figures~\ref{fig:PhaseSpaceMicrostructures}~and~\ref{fig:ModelingParticleDistribution} illustrate the distinct charge modulation that forms under the influence of CSR self-interaction in the micro-bunching instability. Evidently, the modulation pattern is asymmetric as the micro-structures change shape and notably amplitude depending on their position in phase space. Nonetheless, one might be willing to identify an integer number $n_{\mathrm{str}}$ of maxima constituting the charge modulation. Modifying Eq.~(\ref{equ:InstabilityFrequencyEstimate}), we can relate this to the instability frequency
\begin{equation}
f_{\mathrm{inst}} =  n_{\mathrm{str}} \, f_{\mathrm{str}} ~,
\label{equ:InstabilityFrequencyModifiedEstimate}
\end{equation}
where $1/f_{\mathrm{str}}$ denotes the time it takes the micro-structures to perform one full revolution in phase space. As discussed above, this is only loosely related to the oscillation frequency of single particles and may deviate from the nominal synchrotron period.

As reported in~\cite{bol2018}, the integer $n_{\mathrm{str}}$ changes if the shielding by the vacuum pipe is altered by varying its height. Within this subsection, we are interested in examining the corresponding synchrotron motion across these different parameter settings. Figure~\ref{fig:DependenceOnShielding} thus displays the oscillation frequencies of the micro-structures (red) and the single particle frequencies at their positions in phase space (blue). All frequencies are estimated directly above the instability threshold $I_{\mathrm{th}}$, which itself is changing due to the varying shielding~\cite{ban2010}. The dashed red line connects data points with the same number of maxima $n_{\mathrm{str}}$ in the charge modulation pattern, which was estimated by examining the relative charge density $\Delta \psi (q,p,t)$ by eye.

With increased shielding (reduced vacuum gap) the particle frequencies are growing quite smoothly. The oscillation frequencies of the micro-structures, however, show a differing behavior. While they take on very similar values for the vacuum gaps $(\num{19},\num{22},\num{24},\num{27},\num{32}) \, \si{\milli\meter}$, the micro-structure oscillation frequencies are growing much faster with decreasing vacuum gap. This can be observed up until the point where an additional extremum is identified and $n_{\mathrm{str}}$ is incremented. Afterwards, the frequencies of particles and micro-structures are found at very similar values again (e.g.\ transition from $\SI{28}{\milli\meter}$ to $\SI{27}{\milli\meter}$). Figure~\ref{fig:DependenceOnShielding} thereby illustrates again the partial decoupling of the micro-structure propagation in phase space from single particle motion.

The transitions in the observed number of micro-structures and the associated changes in their oscillation frequency are of particular interest as they provide additional insight into the formation process of the micro-structures under different boundary conditions. We consider this a promising starting point for further studies.

\subsection{\label{subsec:AmplitudeAndPositionOfMicroStructures}Amplitude and Position of Micro-Structures}
As explained in subsection~\ref{subsec:HeadTailAsymmetry}, the additional wake potential caused by the micro-structures at the head of the bunch can support and drive the micro-bunching dynamics. Larger local charge modulations lead to a larger perturbation by the additional wake potential which then results in the individual particles being driven to larger oscillation amplitudes. Following this chain of thought, naturally we expect a correlation between the maximum amplitude of the occurring micro-structures and their maximum longitudinal deviation from the origin in phase space.

\begin{figure}[t]
\centering
\includegraphics{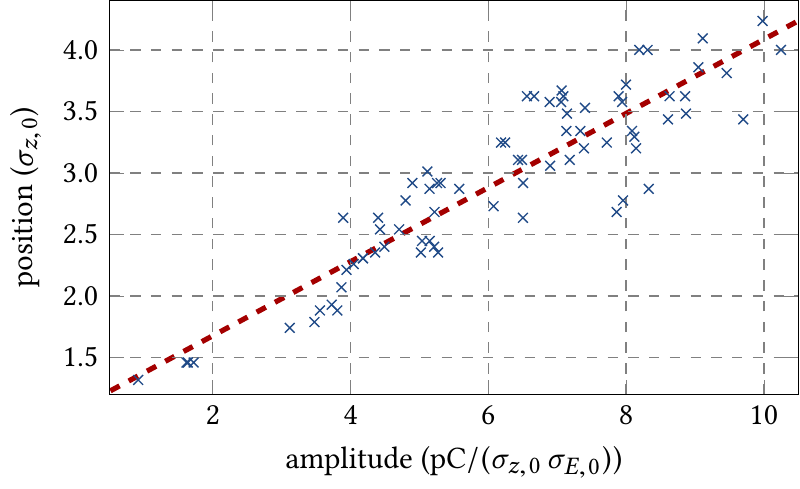}
\caption[Correlation of maximum micro-structure amplitude and maximum position]
{Correlation between the maximum amplitude and maximum longitudinal position of the occurring micro-structures. Shown is simulation data for bunch currents between $\SI{260}{\micro\ampere}$ and $\SI{1000}{\micro\ampere}$. The dashed red line illustrates the clear linear correlation with a correlation coefficient of $\num{0.91}$.}
\label{fig:AmplitudePositionCorrelation}
\end{figure}

In order to verify this, FIG.~\ref{fig:AmplitudePositionCorrelation} displays the maximum amplitude and maximum longitudinal position of the occurring micro-structures for a range of bunch currents between $\SI{260}{\micro\ampere}$ and $\SI{1000}{\micro\ampere}$. With increasing current the strength of the perturbation caused by the CSR self-interaction increases, which leads to larger amplitudes of the micro-structures within the bunch. Figure~\ref{fig:AmplitudePositionCorrelation} illustrates that this corresponds to a larger deviation from the origin on the longitudinal axis as the particles are exposed to a stronger CSR wake potential. For the simulation data considered here, we find a clear linear correlation (as illustrated by the dashed red line) with a correlation coefficient of $\num{0.91}$.

\subsection{\label{subsec:InterpretationOfMeasurements}Interpretation of Measurements}
Finally, we discuss the implications of the previously gained insights on the interpretation of the measurements taken at the KIT storage ring KARA.

\begin{figure}[t]
\centering
\includegraphics{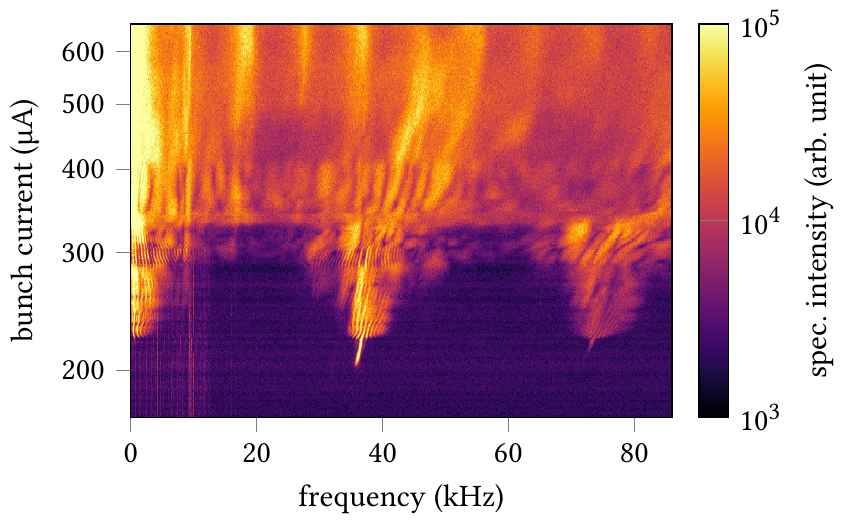}
\caption[CSR Spectrogram Measurement]
{Measurement of the micro-bunching instability at the KIT storage ring KARA (data already published in~\cite{sch2017}). Shown is a CSR power spectrogram in which each horizontal line displays the Fourier transform of the CSR power signal at this particular bunch current. While the specifics of the image may change dependent on machine parameters, some characteristic features are always observed.}
\label{fig:CSRSpectrogramMeasurement}
\end{figure}

One convenient way to observe the micro-bunching dynamics at the storage ring is by measuring the emitted CSR power in the THz frequency range. At KARA, the in-house developed KAPTURE system~\cite{cas2017} allows for a continuous readout of the THz detectors on a turn-by-turn basis. The measurement of this signal over decaying bunch current provides information about the current dependent dynamics of the micro-bunching instability. As an efficient and concise way to display the results, the data is typically post-processed and visualized in CSR power spectrograms~\cite{bro2020diss,ste2019diss} as shown in FIG.~\ref{fig:CSRSpectrogramMeasurement}. Here, each horizontal line displays the Fourier transformed CSR power signal measured at the corresponding bunch current with the color code indicating the spectral intensity. Although the exact details of the image depend on the used machine parameters (e.g.\ momentum compaction factor or acceleration voltage), several features are always present and are considered characteristic for the micro-bunching instability. Right above the instability threshold, here at roughly $\SI{210}{\micro\ampere}$, the emitted CSR power fluctuates with a single distinct frequency introduced as $f_{\mathrm{inst}}$ earlier. For higher currents this dominant frequency increases slightly until it fans out at about $\SI{230}{\micro\ampere}$. Simultaneously, additional contributions in the frequency range between $\SI{0}{\kilo\hertz}$ and $\SI{5}{\kilo\hertz}$ emerge at the left edge of the figure. This general pattern is always observed and indicates the transition from the longitudinal dynamics directly above the instability threshold to the sawtooth bursting, as e.g.\ initially observed at the SLC damping rings~\cite{kre1993}, at higher bunch currents.

In the previous subsections, we discussed how the propagation of the micro-structures in phase space may deviate from the single particle motion. Nonetheless, given the general distribution of particle oscillation frequencies in Figures~\ref{fig:ParticleTrajectoryStatisticsBelowThreshold}~and~\ref{fig:ParticleTrajectoryStatisticsAboveThreshold}, is seems reasonable to assume that also the collectively formed structures propagate faster if they are located further away from the origin in phase space. Based on this hypothesis we can derive a basic reasoning which  explains the occurrence of some characteristic features in the CSR power spectrogram.

The increasing bunch current leads to a stronger perturbation by the CSR wake potential and thus to a larger amplitude of the occurring micro-structures. As established in subsection~\ref{subsec:AmplitudeAndPositionOfMicroStructures}, this corresponds to a drift of the micro-structure position towards larger longitudinal deviations in phase space. Granted that this leads to a faster oscillation of the structures in phase space ($f^\prime_{\mathrm{str}} > f_{\mathrm{str}}$), the instability frequency is, according to Eq.~(\ref{equ:InstabilityFrequencyModifiedEstimate}), increasing as well. While this explains the slight shift of the instability frequency across current, the simple initial dynamics are only observed in a comparably small current range. In FIG.~\ref{fig:CSRSpectrogramMeasurement}, the dominant frequency fans out at roughly $\SI{230}{\micro\ampere}$ marking a clear transition in the occurring dynamics. At these higher bunch currents, the micro-structures reach an amplitude that can no longer be supported by the corresponding CSR wake potential. Reaching this amplitude, the bunch blows up in size and the structures smear out in phase space. As the increased bunch length leads to a reduced perturbation by CSR, the bunch is mainly shrinking due to radiation damping afterwards. Once the bunch length is short enough, the micro-structures emerge again and continuously grow in amplitude. When forming initially, the structures are located close to the origin in phase space and propagate rather slowly, which corresponds to a low instability frequency observed in the CSR power signal. With increasing amplitude the micro-structures are then driven further outside in phase space and accelerate in oscillation frequency. This leads to a sawtooth-shaped burst of CSR emission up until the point where the micro-structure amplitude becomes too large and the cycle starts anew. The spread out instability frequency is thus caused by the varying oscillation frequencies of the micro-structures during their continuous growth in phase space. These considerations motivate a future study that concentrates on a more detailed analysis of the temporal evolution of the instability frequency in both, simulations and measurements. The rather low frequencies at the edge of the figure, however, correspond to the repetition rate of the described bursting cycle and are thus related to the longitudinal damping time as shown in~\cite{bro2018}.

In order to approach control of the instability, we will thus concentrate on time scales comparable to the formation process of the micro-structures, which is governed by the synchrotron period.

\begin{figure*}[t]
\centering
\includegraphics{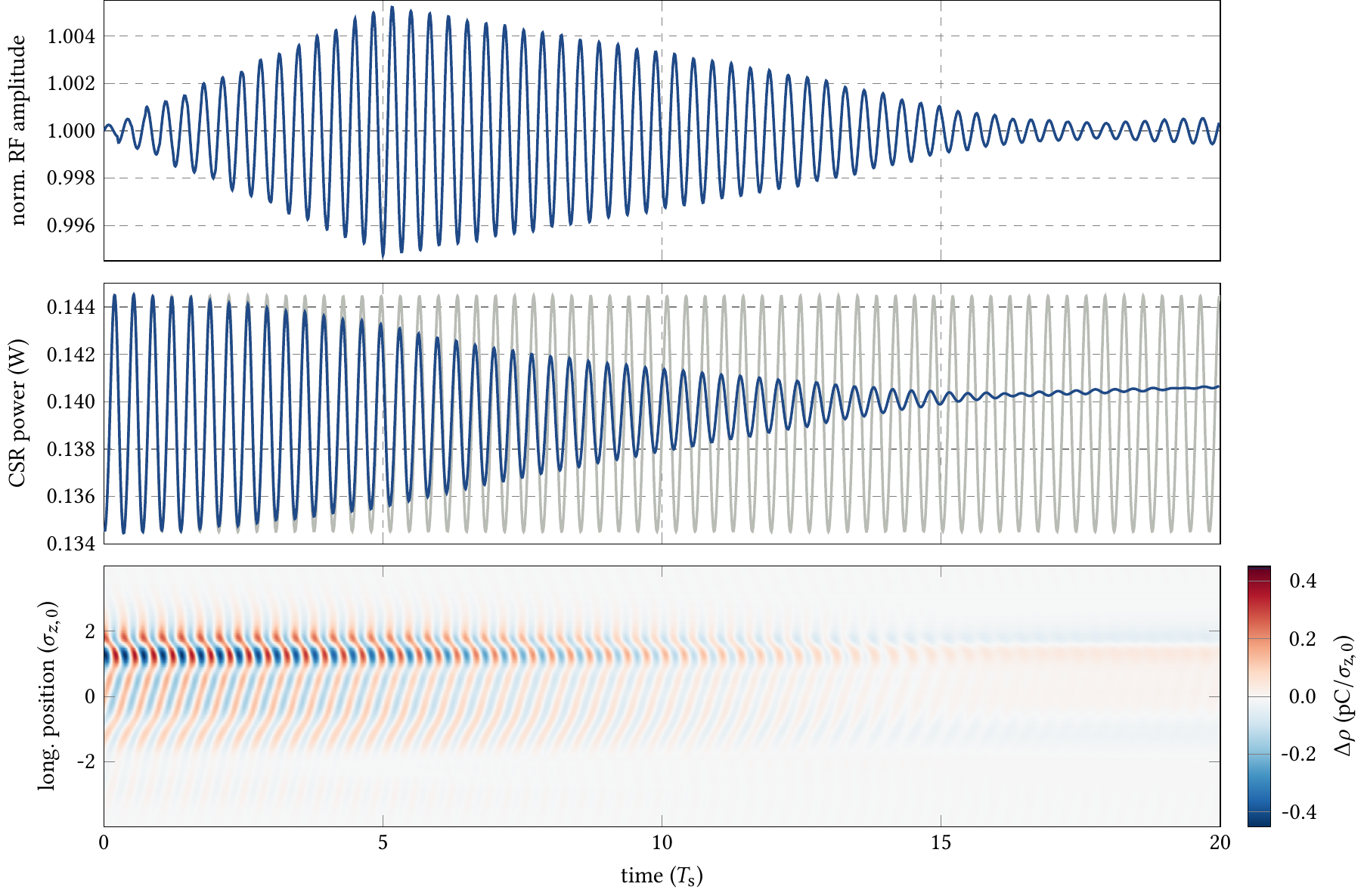}
\caption[Dynamic RF modulation]
{Dynamic modulation of the RF amplitude in order to counteract the perturbation by the CSR wake potential. Careful adjustment of the amplitude and frequency of the RF modulation (top) mitigates the micro-bunching in the longitudinal charge distribution (bottom). This reduces the fluctuation of the emitted CSR power (blue, middle) compared to the natural behavior of the beam (gray, middle). Note that this looks slightly different from FIG.~\ref{fig:ParticleTrajectoriesAndBunchProfiles} as the simulations were done with a lower accelerating voltage.}
\label{fig:DynamicRFModulation}
\end{figure*}

\section{\label{sec:ControlViaDynamicRFModulation}Control via dynamic RF Modulation}
As discussed in the previous section, the micro-bunching dynamics above the instability threshold correspond to a modulation of the restoring force, i.e.\ the slope of the effective potential $V_{\mathrm{eff}}(q)$. That naturally motivates an approach to control of the micro-bunching dynamics by aiming to counteract this perturbation with an RF amplitude modulation
\begin{align}
V_{\mathrm{RF}}(t) &= \hat{V}(t) \sin (2 \pi f_{\mathrm{RF}} \, t) ~, \\
\hat{V}(t) &= \hat{V}_{0} + A_{\mathrm{mod}} \sin (2 \pi f_{\mathrm{mod}} \, t + \varphi_{\mathrm{mod}}) ~.
\label{equ:RFAmplitudeModulation}
\end{align}
While the perturbation by the CSR wake potential cannot be compensated in its entirety, this aims at stabilizing the strength of the restoring force and thereby mitigating the micro-bunching dynamics. By choosing the natural frequency of the occurring instability (e.g.\ observed in the emitted CSR signal) as the modulation frequency $f_{\mathrm{mod}} = f_{\mathrm{inst}}$ and carefully adjusting the amplitude $A_{\mathrm{mod}}$ and phase $\varphi_{\mathrm{mod}}$, the RF modulation can be expected to partially compensate the perturbation by the CSR wake potential. There is however one major complication which has to be considered. Once we start interfering with the natural beam dynamics, the evolution of the longitudinal charge distribution and thus the CSR wake potential is changing as well. Particularly the deliberately chosen modification of the restoring force leads to an altered oscillation frequency and thus significantly affects the synchrotron motion of the present micro-structures. Continuously applying this RF amplitude modulation may therefore initially have the desired effect, but will eventually run out of sync with the perturbation we are aiming to counteract. In that case, the RF modulation may no longer stabilize the restoring force, but actually further drive the instability. Hence, the RF modulation has to be adjusted over time according to the altered micro-bunching dynamics
\begin{align}
A_{\mathrm{mod}} &\rightarrow A_{\mathrm{mod}}(t), \quad f_{\mathrm{mod}} \rightarrow f_{\mathrm{mod}}(t) ~,\\
\hat{V}(t) &= \hat{V}_{0} + A_{\mathrm{mod}}(t) \sin (2 \pi f_{\mathrm{mod}}(t) \, t + \varphi_{\mathrm{mod}}) ~.
\label{equ:DynamicRFAmplitudeModulation}
\end{align}

Figure~\ref{fig:DynamicRFModulation} illustrates the results that can be achieved by applying such a dynamic RF amplitude modulation scheme. The modulation of the RF amplitude (top) clearly results in a mitigation of the micro-structures in the longitudinal charge distribution (bottom) and stabilizes the emitted CSR power (middle). It is worth pointing out that the RF modulation is neither reducing the RF amplitude (on average) nor is it lengthening the bunch. These would both be trivial ways to reduce the strength of CSR self-interaction we are not interested in. In fact, the bunch length is even slightly reduced as indicated by the increase of the CSR power signal towards the end of the displayed time period ($\num{15} \, T_\mathrm{s} < t < \num{20} \, T_\mathrm{s}$). In order to achieve these results the modulation amplitude $A_{\mathrm{mod}} \in \left[ 0, 4.5 \right] \num{e-4} \, \hat{V}_{0}$ and frequency $f_{\mathrm{mod}} \in \left[ 0.233, 0.245 \right] \, f_{\mathrm{s},0}$ were both adjusted at a step width of $\Delta t = \num{0.25} \, T_{\mathrm{s}}$. While this step width might possibly be relaxed to larger values, it has to be chosen small enough to be able to react to the altered dynamics of the micro-bunching. As the synchrotron period $T_{\mathrm{s}}$ governs the time scale of these dynamics, the step width $\Delta t$ has to be chosen in the same order of magnitude.

For the results shown in FIG.~\ref{fig:DynamicRFModulation}, the sequence of modulation amplitudes $A_{\mathrm{mod}}(t_{i})$ and frequencies $f_{\mathrm{mod}}(t_{i})$ was found via empirical testing. However in general, finding this dynamic sequence of RF amplitude modulations is a non-trivial task that depends on the bunch current and the specific machine parameters of the storage ring. As illustrated in~\cite{bol2019}, we can view this as a sequential decision problem where the amplitudes $A_{\mathrm{mod}}(t_{i})$ and frequencies $f_{\mathrm{mod}}(t_{i})$ need to be chosen iteratively, one step after another. Informing these decisions with diagnostics about the current state of the micro-bunching yields a closed feedback loop and provides the basis for the reinforcement learning approach discussed in~\cite{bol2019}.

Finally, it should be mentioned that the bunch currents considered here are mostly below or directly above the instability threshold. While the characteristic sawtooth-like bursting only occurs at higher bunch currents, this behavior is still caused by very similar micro-structure dynamics within the bunch (as discussed in~\cite{bol2017}). Controlling these micro-bunching dynamics should thus imply control over the instability in the sawtooth regime as well.

\section{\label{sec:Summary}Summary and Outlook}
In order to better understand the longitudinal dynamics underlying the micro-bunching instability at storage rings we discussed different aspects of the synchrotron motion under CSR self-interaction. To develop some intuition, we focused on single particle motion and its perturbation by the CSR wake potential. We showed that this system, to good approximation, can be viewed as harmonic oscillator with perturbed linear restoring force. Using this simple model we predicted a quadrupole-like charge modulation below the instability threshold, which was verified using Inovesa simulations. Furthermore, we considered synchrotron motion above the instability threshold focusing on how the single particle motion relates to the formation and propagation of the occurring micro-structures in phase space. We found that the observed charge modulation is not merely caused by resonant motion of single particles, but rather is a collective effect of many particles traveling on varying trajectories. Moreover, structures are formed at the head of the bunch and are located further away from the synchronous position the larger their amplitudes. Eventually, we elaborated on how these insights help interpret measurements taken at KARA and motivated an approach to control of the micro-bunching dynamics.

Following the outlined concept of a longitudinal feedback loop, we propose a method to control the micro-bunching dynamics to an extent that can enable new operation modes at modern storage rings. If the longitudinal charge distribution can be kept stable at these small bunch lengths, it would significantly extend the limits of the longitudinal beam properties that can be provided to experiments. Given the highly developed diagnostics at KARA, the storage ring seems particularly well-suited for these efforts and first studies are ongoing.

\section{\label{sec:Acknowledgements}Acknowledgements}
T.\ Boltz and P.\ Schreiber acknowledge the support by the DFG-funded Doctoral School \enquote{Karlsruhe School of Elementary Particle and Astroparticle Physics: Science and Technology (KSETA)}.

\bibliography{paper}
\end{document}